\documentclass[12pt]{article}
\pdfoutput=1
\usepackage{jcapmod}
\usepackage{color}
\usepackage{colordvi}
\usepackage{setspace}
\usepackage{graphicx}
\usepackage{bm}
\usepackage{hyperref}
\usepackage{epsfig}
\usepackage{booktabs}
\usepackage[english]{babel}
\usepackage{amsmath,amssymb,amsbsy,amstext, amsthm, sim
plewick}
\usepackage{hyperref}
\usepackage{graphicx}
\usepackage{amsfonts}
\usepackage{amssymb}
\usepackage{upgreek}
\usepackage{simplewick}
 \usepackage{exscale,relsize}

\usepackage[margin=.8cm]{caption}

\usepackage{colortbl}
\definecolor{lightgreen}{cmyk}{0.2, 0, 0.2, 0.2}
\definecolor{lightgray}{cmyk}{0.1,0.2,0,0.1}
\definecolor{lightgray2}{cmyk}{0.1,0.1,0,0.1}

\makeatletter
\newlength{\apb@width}
\newcommand{\autoparbox}[2][c]{\settowidth{\apb@width}{#2}\parbox[#1]{\apb@width}{#2}}

\makeatother

\numberwithin{equation}{section}

\def\beq{\begin{equation}}
\def\eeq{\end{equation}}

\def\bea{\begin{eqnarray}}
\def\eea{\end{eqnarray}}

\def\beq{\begin{equation}}
\def\eeq{\end{equation}}
\def\bea{\begin{eqnarray}}
\def\eea{\end{eqnarray}}

\def\Mp{M_{\rm pl}}

\def\0{{\vec{0}}}
\def\k{{\vec{k}}}
\def\q{{\vec{q}}}
\def\p{{\vec{p}}}

\DeclareRobustCommand{\SkipTocEntry}[4]{}

\def\Mp{M_{\rm pl}}

\DeclareSymbolFont{extraup}{U}{zavm}{m}{n}
\DeclareMathSymbol{\varheart}{\mathalpha}{extraup}{86}
\DeclareMathSymbol{\vardiamond}{\mathalpha}{extraup}{87}

\begin{document}

\begin{titlepage}

\setcounter{page}{1} \baselineskip=15.5pt \thispagestyle{empty}
\bigskip\
\bigskip\
\begin{center}
{\fontsize{24}{24}\selectfont  \sc On Squeezed Limits\\ in Single-Field Inflation \\[.7cm] Part I}
\end{center}
\vspace{.2cm}

\begin{center}
{\fontsize{13}{30}\selectfont  Raphael Flauger$^{1, 2}$, Daniel Green$^{1,3}$, and Rafael A. Porto$^{1,4}$}
\end{center}

\begin{center}
\vskip 8pt
\textsl{$^1$ School of Natural Sciences,
 Institute for Advanced Study,
Princeton, NJ 08540, USA}

\vskip 5pt
\textsl{$^2$  Center for Cosmology and Particle Physics, Department of Physics,
New York University, New York, NY, 10003, USA}

\vskip 5pt
\textsl{$^3$ Kavli Institute for Particle Astrophysics and Cosmology \&  Stanford Institute for Theoretical Physics,  Stanford, CA 94306, USA}

\vskip 5pt
\textsl{$^4$ Department of Physics \& ISCAP, Columbia University, New York, NY 10027, USA }

\end{center}

\vspace{.2cm}
{ \noindent \textbf{Abstract} \\
\hrule \vspace{0.3cm}
\noindent The $n$-point correlation functions in single-field inflation obey a set of consistency conditions in the exact squeezed limit which are not present in multi-field models, and thus are powerful tools to distinguish between the two. However, these consistency conditions may be violated for a finite range of scales in single-field models, for example by departures from the Bunch-Davies state. These {\it excited} states may be the consequence of interactions during inflation, or may be a remnant of the era that preceded inflation.  In this paper we analyze the bispectrum, and show that in the regime of theoretical control the resulting signal in the squeezed limit remains undetectably small in all known models which continuously excite the state. We also show that the signal remains undetectably small if the initial state is related to the Bunch-Davies state by a Bogoliubov transformation and the energy density in the state is small enough so that the usual slow-roll conditions are obeyed. Bogoliubov states that lead to violations of the slow-roll conditions, as well as more general excited states, require more careful treatment and will be discussed in a separate publication.

}
 \vspace{0.3cm}
 \hrule

\vspace{0.6cm}
\end{titlepage}

\tableofcontents

\newpage
\section{Introduction}

Observations of the cosmic microwave background (CMB) and large scale structure (LSS) provide compelling evidence for nearly Gaussian, nearly scale-invariant primordial perturbations, just as one might have expected from a phase of inflation at very early times \cite{wmap,spt,act,lss}. With the case for an early inflationary phase strengthening, it is timely to ask to what extent we will be able to answer more detailed questions about the very mechanism that produces it. One question that has gathered significant attention in recent years, because it is both interesting and seems answerable, is whether there was a single active light degree of freedom during this phase or many. 

An observable that is very sensitive to the spectrum of excitations during inflation is the squeezed limit of the three-point function of curvature perturbations. It has been shown to be of the form\footnote{As usual $k_S$ and $k_L$ stand for the short and long wavelength modes, $P_\zeta$ is the power spectrum without the $\delta$-function, and $\langle \cdots \rangle '$ means we dropped the overall $(2\pi)^3 \delta^3 (\k_1 + \k_2 +\k_3+\cdots)$.}
\beq\label{consist}
\lim_{k_L\to 0} \langle\zeta_{k_L}\zeta_{k_S}\zeta_{k_S}\rangle'=-P_{\zeta}(k_L)P_{\zeta}(k_S)\left[(n_s-1) + O\left(k_L^2/ k_S^2\right)\right],\eeq
for models with a single light field \cite{Maldacena:2002vr,Creminelli:2004yq,consist2,cremi,consist3,consist4,consist5}, and more generally in dissipative models with multiple light fields but a single {\it clock}~\cite{diseft,diseft2}.

The expression in Eq.~(\ref{consist}), known as the `consistency condition', is an expansion in powers of $k_L/k_S$ which is valid to all orders in the slow-roll parameters. Its meaning is most transparent in the limit in which the wave-number vanishes, i.e. $k_L \simeq 0$. In this limit, Eq.~(\ref{consist}) states (in co-moving coordinates) that the local physics is unaffected by the presence of long-wavelength modes. The fact that the leading corrections for small but finite values of $k_L$ scale as $(k_L/k_S)^2$ is a manifestation of the equivalence principle. In other words, we measure deviations from  free-fall through curvature effects. 

At first sight, the consistency condition appears very useful in distinguishing between different models of inflation. Any three-point function that does not behave as in Eq.~(\ref{consist}) when $k_L/k_S \to 0$ cannot derive from single-clock models. In practice, however, when we constrain models using data, we are forced to content ourselves with a finite range of momenta. 
For single-clock models, the behavior in Eq.~(\ref{consist}) will eventually take over as $k_L/k_S \to 0$, but it may be violated for the finite $k_L/k_S$ accessible to a given experiment. In order to rule out single-clock inflation experimentally, we must thus establish whether such models can produce a detectable non-trivial scaling for the three-point function for the observable range of momenta. Put differently, one needs to understand for which values of $k_L/k_S$ single-clock models allow for deviations from Eq.~(\ref{consist}).

Scenarios that lead to a non-trivial behavior in the squeezed limit must involve either non-attractor dynamics~\cite{Namjoo:2012aa,Chen:2013aj}, or departures from the Bunch-Davies state. We will focus on the latter. There are two possible roads to excited states during inflation: either the state becomes excited by violations of adiabaticity in time-dependent processes during inflation, or one assumes that inflation started in an excited state. If we assume that the excited initial state was generated by some unspecified but causal pre-inflationary dynamics, both scenarios lead to non-trivial correlations between long and short wavelength modes when all the modes are inside the horizon. In this regime, one can use flat-space intuition to explain why single-clock models require an excited state to generate a non-trivial squeezed limit. Intuitively, since one can show that correlations in the Bunch-Davies state from early times are exponentially suppressed, excited states provide the necessary energy to produce non-trivial correlations that survive at late times. We will make this argument more precise throughout the paper.

Whether obtained dynamically or postulated as an initial condition, the additional energy density carried by the excited state will not only be responsible for the correlations, but will also affect the overall dynamics of the system. If it becomes too large, it may lead to a breakdown of the assumptions entering in our calculations, such as a meaningful separation into background and perturbations, or disrupt the inflationary dynamics altogether. Furthermore, the non-adiabatic time-dependent processes themselves must be sufficiently weakly coupled (since our universe is almost Gaussian). Otherwise very little can be said about the resulting evolution of the curvature perturbations, and in particular whether there is a non-trivial scaling for the three-point function. 

In this paper we will take these considerations seriously and show that both for models in which the state is excited by non-adiabatic evolution, and for models with Bogoliubov initial states with significant occupation numbers but with energy densities small enough not to spoil the slow-roll conditions, non-trivial correlations can only exist between modes that are separated in their wave numbers by less than the fourth root of the observed power spectrum, or approximately two orders of magnitude, \beq \label{bkskl} \frac{k_L}{  k_S } \lesssim \Delta_\zeta^{\frac12}\sim 10^{-2}.\eeq 
We will discuss the consequences of this bound in detail. In summary, for models in which the state was excited by non-adiabatic evolution, it implies that they cannot produce a signal in current or planned experiments that are sensitive predominantly to the squeezed limit of the bispectrum. For the Bogoliubov states with large occupation numbers the bound (\ref{bkskl}) implies that they cannot lead to a power spectrum that is scale-invariant over more than two orders of magnitude. However, from measurements of the power spectrum from the microwave background, galaxy surveys, and the Ly$\alpha$ forest, we know that the primordial power spectrum is nearly scale invariant over at least three orders of magnitude~\cite{Hlozek:2011pc}, thus ruling out Bogoliubov initial states with large occupation numbers.

For Bogoliubov initial states with small occupation numbers and energy densities below the kinetic energy density of the inflaton, there are no constraints from the power spectrum, and the bound (\ref{bkskl}) is also weakened. We can thus not rule them out immediately. However, we will provide signal-to-noise estimates for Planck and Euclid showing that these models do not lead to observational signatures. 

Once we allow the energy density stored in the excited state to exceed the kinetic energy of the inflaton, there are several things that require additional care. First, the energy density in the excited state is now large enough to affect the background evolution. In particular, it determines the evolution of $\dot{H}$ and since the energy density in the excitations redshifts like radiation, we now have a phase with $\delta\approx -2$. Second, on small scales $k/a>\dot\phi^{1/2}$ the so-called $\zeta$-gauge is no longer well-defined. This is not too alarming, and of course also happens in the present-day universe, but requires some additional care.  For this reason, we will limit ourselves to states for which the energy density stored in the excited state is negligible compared to the kinetic energy of the inflaton in the present paper and postpone a careful study of the regime in which this condition is not met to~\cite{future}. It should be noted that, strictly speaking, previous studies of excited initial states are also limited to this regime.

States that are related to the Bunch-Davies state by a Bogoliubov transformation are superpositions of states with an even numbers of particles. As a consequence, the three-point correlations have to be generated through interactions, which are small in the case of slow-roll inflation. General initial states may have non-trivial three-point correlations even in the absence of interactions, and we will also consider them in~\cite{future}.

In section~\ref{sec:orig}, we illustrate both how the consistency relation can be violated for a finite range of momenta and what prevents us from making this range arbitrarily large. We use the lessons from this section to derive the constraint~\eqref{bkskl} for models in which the state is excited through non-adiabatic evolution as well as for models with Bogoliubov initial states with significant occupation numbers but with energy densities below the kinetic energy density of the inflaton in section~\ref{sec:slow1}. 
In Section~\ref{sec:eft} we generalize our results using the effective field theory (EFT) of inflation~\cite{Creminelli:2006xe, Cheung:2007st}. However, the essence of the more general argument will be captured in sections \ref{sec:orig} and \ref{sec:slow1}. The more phenomenologically oriented reader, if satisfied with the exposition and results in these sections, may thus proceed directly to section~\ref{sec:obs} for an account of observational perspectives.  We conclude in section~\ref{sec:disc}.

\section{Origin of a non-trivial squeezed limit}\label{sec:orig}

To understand what is required for a non-trivial behavior in the squeezed limit, let us first give a more intuitive argument for the validity of the consistency relation in the exact squeezed limit. Once the attractor has been reached, in single clock inflation modes freeze out after they exit the horizon. A frozen mode that is larger than the horizon is indistinguishable from a rescaling of the background and is therefore not measurable inside the horizon. As a consequence, the power spectrum of the short modes will not be modified at horizon crossing in the presence of the long wavelength mode. Therefore, if all correlation functions are determined when the shortest wavelength mode crosses the horizon, the consistency consistency will be obeyed.  
 
To evade the consistency condition in single-clock inflation, a correlation between long and short modes must have been generated before the long wavelength mode crossed the horizon. In the exact squeezed limit, i.e. for $k_L \simeq 0$ this cannot happen because the long wavelength mode is {\it always} outside the horizon. This is the limit in which the consistency condition applies. However, for finite values of momenta it need not be the case. Before horizon crossing, the long mode is distinguishable from the background and can induce correlations with the short wavelength modes.  This correlation acts as an initial condition for modes observed a late times, and therefore it is observable.

It is well-known \cite{Maldacena:2002vr, consist2} that the three-point function computed in the Bunch-Davies state does not depart from the scaling in Eq. (\ref{consist}). What prevents such correlations from occurring?
\vskip 6pt

The answer is that, in the Bunch-Davies state, the contributions to the bispectrum from modes that are deep inside the horizon ($k/a \gg H$) are exponentially suppressed.  We can understand the origin of this suppression from flat space intuition since the modes inside the horizon are essentially insensitive to the curvature.  For illustrative purposes, let us then consider $in$-$in$ correlation functions in a time independent system.  For example, if we have a scalar field $\varphi$ with an interaction Hamiltonian $H_{\rm int} = \tfrac{1}{6} \mu \varphi^3$, the equal time correlation three-point function is given by
\bea\label{phiflat}
\langle \varphi_{\k_1} \varphi_{\k_2} \varphi_{\k_3} (t) \rangle' &= & -i \tfrac{1}{8 k_1 k_2 k_3} \int_{-\infty(1-i\epsilon)}^t dt' \mu e^{ i(k_1 +k_2 + k_3)(t'-t)} - {\rm c.c.} \\
&=&  -\mu\frac{1}{4 k_1 k_2 k_3(k_1 +k_2 +k_3)}\nonumber\ , 
\eea
where $k_i=|\vec{k}_i|$ is the energy of the mode with momentum $\k_i$.  (Note the parallelism with the computation in an expanding universe, in particular the factor of $1/k_t$, with $k_t \equiv k_1+k_2+k_3$. The main difference, of course, is the late time behavior of the correlators in de Sitter, where curvature effects start playing a role.) Although it appears as if the correlation function receives contributions from all times, the contribution from early times is exponentially suppressed due to the rapid oscillations.

\begin{figure}[t!] %  figure placement: here, top, bottom, or page
 \begin{center}
 \includegraphics{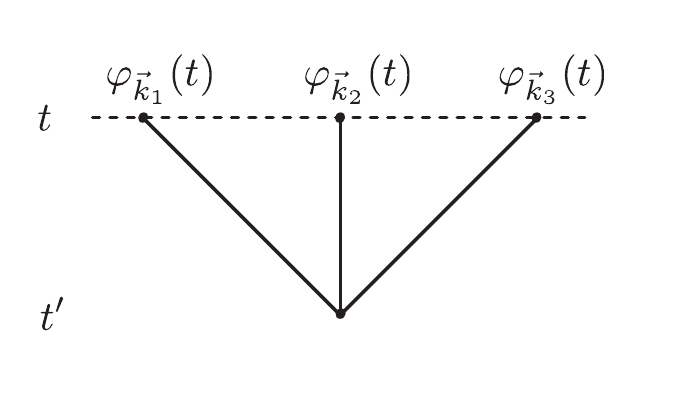}
 \hspace{1cm}
 \includegraphics{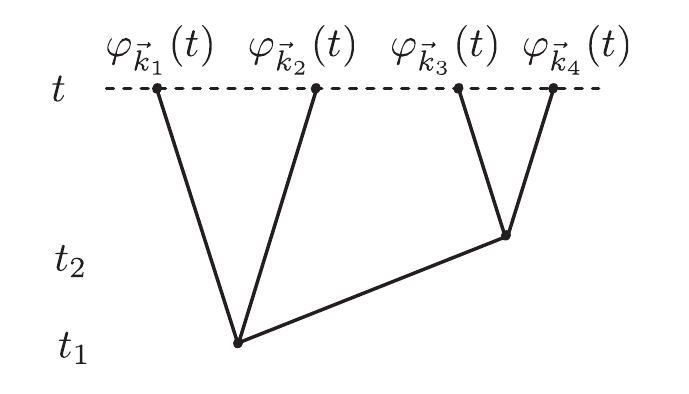}
 \caption{Sample processes responsible for correlations in the three- and four-point function given in equations~\eqref{phiflat} and~\eqref{phiflat2}}
 \label{fig1}
 \end{center}
\end{figure}

One can intuitively understand the origin of the correlations, and the fact that contributions from early times are exponentially suppressed, as follows. The computation that leads to the result in~\eqref{phiflat} represents the production of three {\it virtual} quanta from the vacuum at some point in spacetime that are subsequently ``measured'' by the field(s) $\varphi_{\k_i}(t)$. We describe this process diagramatically in the left panel of Fig.~\ref{fig1}. 
Since all three quanta originate from the same event, they are correlated. Because they carry energy, energy conservation prevents them from going on-shell, implying that they could not have been produced long before they are measured at time $t$. However, energy conservation may be violated quantum mechanically for a period of time $\Delta t\sim1/\Delta E=1/k_t$. We conclude that the quanta must have been created less than $\Delta t$ prior to $t$, explaining why the contributions to the integral from early times are exponentially small. This intuition implies that also in the Bunch-Davies state correlations are generated near $t' \simeq t-\Delta t$. When translated to $\zeta$, this implies that the correlations are generated shortly before the short mode freezes. Since the long mode is already frozen at this time, this then ultimately turns into the consistency condition given in Eq. (\ref{consist}).

The reader might complain that the case of the three-point function is not very generic, because there is only one possible way these quanta can be created. To address this, let us consider the four-point function $\langle \varphi_{\k_1} \varphi_{\k_2} \varphi_{\k_3} \varphi_{\k_4} (t) \rangle'$ as a less trivial example. One of the processes that is responsible for these correlations is represented in the right panel of Fig.~\ref{fig1}. Three virtual quanta are created from the vacuum with momentum $(\k_1,\k_2,\q\equiv-\k_1-\k_2)$ at time $t_1$. Two of them are measured by two of the fields at time $t$, while the third decays at a time $t_2\simeq t_1+\Delta t_1$ (with $\Delta t_1 \simeq 1/(k_1+k_2+q)$ in agreement with the uncertainty relation) into two quanta with momentum $(\k_3,\k_4)$ which are subsequently measured also at time $t$ by two other fields. Since the presence of the four quanta violates energy conservation by $\Delta E=k_1+k_2+k_3+k_4\equiv k_t$, the instant $t_2$ should be within $\Delta t=1/k_t$ of the instant $t$ at which they are measured. This process thus results in the contribution to four-point function given in the first line below
\bea
\label{phiflat2}
\langle \varphi_{\k_1} \varphi_{\k_2} \varphi_{\k_3} \varphi_{\k_4} (t) \rangle' 
&=&  \mu^2\frac{1}{16k_1 k_2 k_3 k_4} \frac{1}{q}\left[\frac{1}{(k_1+k_2+q)} \frac{1}{k_t}\right.\\&&\left.+\frac{1}{(k_3+k_4+q)} \frac{1}{k_t}+\frac{1}{(k_1+k_2+q)} \frac{1}{(k_3+k_4+q)}\right] + {\rm permutations}\nonumber\,.
\eea 
The second line arises from similar processes. We see that also for higher $n$-point functions correlations generated at early times are exponentially small, simply because quanta had to originate from the vacuum state and energy conservation prevents them from going on-shell. The more energetic the quanta the closer to the measurement at time $t$ the correlations are produced. This also implies that these correlation functions do not exhibit non-trivial behavior in the limit as one of the external momenta becomes soft. Note, however, that there is a pole in this correlation function as $q\to0$ in the internal line (because the particles are massless). This pole ultimately leads to the consistency condition for the four-point function in the collinear limit. (See \cite{Leblond:2010yq,Senatore:2012wy,paolo,lam,dd4,KehaRiot} for more details on soft and collinear limits.)\\

As we emphasized, the consistency relation must hold once the long wavelength mode is outside the horizon. Therefore any violation must arise while all the modes are inside the horizon and our flat space intuition (and notion of energy conservation and uncertainty principle) applies. In what follows we will then use this terminology loosely while referring to the possible realizations of a non-trivial scaling in the squeezed limit. There were two important assumptions that lead to the exponential suppression: a) the system is time translation invariant, and b) we are computing vacuum correlation functions.  If either one of these assumptions is violated, then the correlations are not necessarily suppressed. This is because energy conservation does not prevent the configuration from being realized. In the context of inflation, we can easily break either of these assumptions. In our flat space example, we might imagine an interactions of the sort $\mu(t)\varphi^3$, with $\mu(t) = \cos(k_\star t)$ arising the time evolving background field $\phi(t)$. Now the contributions to the integral are not suppressed when $k_t = k_\star$.  This is consistent with energy conservation, as we have extracted energy $k_\star$ from the background $\phi(t)$ to produce real quanta. In this simple example, the reader may recognize the possibility of {\it resonances} and oscillations in the power spectrum.

The second assumption, that we are in the vacuum, may also be violated during or prior to inflation. Although the energy density will redshift away during expansion,  at the onset of inflation it is plausible that the fluctuations were in an excited (i.e. finite energy) state. Energy conservation no longer forbids contributions to the correlations from early times because we are starting in a finite energy state, again, given by some scale $k_\star$. In the flat space example, if the mode $k_1$ is excited, a negative frequency $-k_1$ can appear in the integral as $e^{i \tilde k t}$,  with $\tilde k \equiv -k_1 +k_2 +k_3$. Now we get contributions at all times when $\tilde k \simeq 0$, namely $k_1=k_2+k_3$. The reader may already recognize this as the contribution from excited states to the so-called {\it folded} shape.

An important feature of both the exceptions we mentioned above, is that they are associated with some positive energy scale, i.e. $k_\star$. For the case of departures from the vacuum state, there will be an energy density present from the beginning of inflation which will back react on the geometry, and therefore it must be bounded in order to retain the basic features of the inflationary dynamics. On the other hand, for the case of resonances, the scale $k_\star$ cannot be too large or else interactions induced by the non-adiabatic evolution may become strongly coupled, in which case very little can be said about the resulting evolution of the curvature perturbations, and in particular whether or not a non-trivial scaling develops. 
We then conclude that the necessary condition $k_t \lesssim k_\star$ for a non-trivial scaling in the squeezed limit must itself be bounded.  We will show how these considerations lead to an universal bound on $k_\star$, hence to a lower bound on $k_L/k_S$.

One might ask about the validity of the consistency relation when correlations are introduced as initial conditions before inflation. In this paper we will assume causal pre-inflationary dynamics implying that modes inside the horizon $H(t_0)$ are uncorrelated with modes outside the horizon at that time. This implies the validity of the consistency condition when $k_L/k_S < a(t_0)H(t_0)/k_\star$. 

\section{Bounds on squeezed limits in slow-roll inflation}\label{sec:slow1}

Let us now derive a bound on the range of wave numbers that can be correlated in a non-trivial way for inflationary models that (at least on average) satisfy the usual slow-roll conditions 
\bea
\epsilon=-\frac{\dot{H}}{H^2}\ll1\qquad\text{and}\qquad \delta=\frac{\ddot{H}}{2 H\dot{H}} \ll 1\,.
\eea
We will first discuss models in which the state is being excited continuously during inflation and then turn to slow-roll models in which the state was excited at some fixed early time by some unspecified dynamics (plausibly one which violates the slow-roll conditions). The discussion of states that are sufficiently excited so that at least for some period of time $\delta\approx -2$ will be postponed to~\cite{future}.

As we already explained, correlations between a long mode with wave number $k_L$ and two short modes with wave numbers $k_S$ must have been generated before (or around) $k_L \sim a(t) H$ when $k_S \gg a(t) H$.  We also know, from energy conservation, that at any given time only modes with wave numbers up to $k_S \lesssim k_\star$ are significantly excited and can contribute to non-trivial correlations at late times. Our goal will be to derive an upper bound on $k_\star$, hence on $k_S/k_L$. In the case of non-adiabatic evolution, it will correspond to the validity of a weak coupling expansion for the perturbations. In the case of the excited initial state, it will follow from the validity of the slowly-varying background evolution conditions. In either case it turns out to be of the form \[k_\star/a \lesssim \dot\phi^{1/2}\,.\] Therefore, if modes are significantly excited, non-trivial correlations can only arise between modes separated by
%\beq \frac{k_S}{k_L} \lesssim \dot\phi^{\tfrac{1}{2}}/ H,~~\left( {\rm or}~~ \frac{k_L}{k_S} \gtrsim H/\dot\phi^{\tfrac{1}{2}}\right).
%\eeq
\beq \frac{k_L}{k_S} \gtrsim H/\dot\phi^{1/2}\,.
\eeq
We recognize the right hand side as the fourth root of the power spectrum\footnote{Here $\Delta_\zeta^2=2\pi^2\Delta_\mathcal{R}^2$, where $\Delta_\mathcal{R}^2\approx 2.4\times 10^{-9}$~\cite{wmap}. }
\beq
k^3 \langle \zeta_\k \zeta_{-\k}\rangle' \equiv \Delta^2_\zeta = \frac{H^4}{2 \dot \phi^2} \approx 4.7\times10^{-8}, \,
\eeq
which allows us to re-write our bound as
\beq
\frac{k_L}{k_S} \gtrsim \frac{H}{\dot \phi^{\tfrac{1}{2}}} \simeq \Delta_\zeta^{1/2} \simeq 10^{-2}.
\eeq
The above expression for the power spectrum as written is only valid for the Bunch-Davies state, but we will show in what follows that in essence our simple line of arguments remains unchanged.

As a result, COBE normalization puts an upper bound on the range of wave numbers $k_S/k_L$ that can display non-trivial behavior in the squeezed limit. \vskip 6pt

\subsection{Non-adiabatic evolution}\label{nonadia}
 
Let us now consider the case in which non-trivial correlations are generated by interactions during inflation in more detail.

In this case, the Hamiltonian typically has a significant time-dependence on a time scale $\Delta t \simeq \omega_\star^{-1} \ll H^{-1}$. As long as the natural frequency of the modes is larger than this frequency, $k/a>\omega_\star$, they evolve adiabatically and remain in the Bunch-Davies state. However, as they redshift their frequency eventually becomes comparable to $\omega_\star$ and the state becomes excited, leading to non-trivial correlations. Concrete examples that we now discuss are resonant models \cite{Chen:2008wn,Flauger:2009ab,Flauger:2010ja} or dissipative models \cite{diseft, warm,trapped,unwind}.

\subsubsection{Resonances}

As a first concrete example, let us consider a model of single-field inflation with the potential \cite{Flauger:2009ab,Flauger:2010ja}
\beq\label{equ:resint0}
 V = V_0 + \Lambda^4 \cos\left(\frac{\phi}{f}\right),
\eeq
where $V_0$ is a potential that supports slow-roll inflation, and $\Lambda, f$ are energy scales. As the inflaton slowly rolls over the oscillatory potential, the Friedmann equation implies that the background geometry will oscillate with frequency $\omega_\star \simeq \dot{\phi}/f$.\footnote{Note that for a slowly rolling $\phi$ this is reminiscent of the time-dependent coupling we discussed in flat space, i.e. $\mu(t) \simeq \cos(\omega_\star t)$.}
We assume that the modes start out in the Bunch-Davies state at early times when $k/a\gg\omega_\star\gg H$. As the modes redshift their natural frequency becomes comparable to the frequency of the background and they eventually undergo parametric resonance and the state becomes excited. This intuition is in perfect agreement with the solution of the curvature perturbation
\bea
\zeta_{\vec{k}}(t)=\zeta_k(t)a(\vec{k})+\zeta_k^*(t)a^\dagger(-\vec{k})\,.
\eea
Up to slow-roll suppressed corrections, the mode function after the resonance is given by
\bea
\zeta_k(t)=\zeta_k^{(o)}\left[i\sqrt{\frac{\pi}{2}}x^{3/2}H^{(1)}_{3/2}(x)-i \beta_k \sqrt{\frac{\pi}{2}}x^{3/2}H^{(2)}_{3/2}(x)\right]\,,
\eea
where $x=k/aH$. As anticipated, the non-zero Bogoliubov coefficient $\beta_k$
\bea\label{eq:betares}
\beta_k=3b_0\sqrt{\frac{\pi}{2}}\alpha^{-1/2}e^{-i\frac{\phi_k}{f}}\qquad\text{with}\qquad \alpha=\frac{\omega_\star}{H}\qquad\text{and}\qquad b_0 = \frac{\Lambda^4}{ V_0'(\phi_0) f}\,,
\eea
shows that the resonance has lead to an excited state. Here $\phi_0$ and $\phi_k$ denote the value of the scalar field when the pivot scale $k_0$ and the mode with comoving momentum $k$ exit the horizon respectively.
Since $|\beta_k|\ll1$, the normalization of the power spectrum is thus still given by the slow-roll result
\beq
\Delta^2_\zeta = \frac{H^4}{2 \dot \phi^2} \simeq 4.7\times 10^{-8} \ .
\eeq
Formula~\eqref{eq:betares} also reveals that the phase of $\beta_k$ is oscillatory as a function of $k$ (rather than constant, which is often assumed for excited initial states), which will be relevant to the observational consequences discussed in section \ref{sec:obs}.

The bispectrum in the resonance model receives its largest contribution from the interaction~\cite{Chen:2008wn}
\beq
H_{\rm int} = - \int d^3x dt a^3(t) \epsilon(t) \dot \delta(t) \zeta^2 \dot \zeta\,.
\eeq
It gives rise to \cite{Flauger:2010ja}
\bea\label{equ:resbi}
&&\langle \zeta_{\k_1} \zeta_{\k_2} \zeta_{\k_3} \rangle' = 4 \Delta_\zeta^4  \frac{f^{\rm res}}{k_1^2 k_2^2 k_3^2} \Big[ \sin\Big( \alpha \log (\tfrac{k_1+k_2+k_3}{k_0}) \Big)\nonumber \\ &&\hskip 5.2cm+\frac{1}{\alpha} \sum_{i\neq j} \frac{k_i}{k_j} \cos\Big( \alpha \log (\tfrac{k_1+k_2+k_3}{k_0}) \Big)+ {\cal O}(\alpha^{-2}) \Big]\,,
\eea
where the amplitude is given in terms of the underlying parameters of the model as 
\beq
f^{\rm res} = \frac{3 b_0 \sqrt{2\pi}}{8} \alpha^{3/2}\,.
\eeq
To see that there is indeed an intermediate regime in which the consistency condition is violated, let us first consider the behavior of the bispectrum for squeezed configurations in which $k_1\ll k_2,k_3$ satisfying $\alpha^{-1} < \frac{k_L}{k_S} \ll 1$ with $\k_L=\k_1$ and $\k_S=(\k_2-\k_3)/2$. In this regime the term on the first line dominates over the term on the second line. Not also that in this regime we cannot Taylor expand the sine in (\ref{equ:resbi}) in a series in $k_L$ because $\alpha \frac{k_L}{k_S} \gtrsim 1$.  As a result, we find $B(k_L,k_S ,k_S) \propto 1/k_L^2$ which violates the consistency condition.  Now consider configurations with $\frac{k_L}{k_S} \ll \alpha^{-1}$.  For these modes the second term dominates and we can expand the oscillatory term. We find that in this regime the consistency condition holds~\cite{Flauger:2010ja}. By going to higher orders one finds that the leading corrections go as $(k_L/k_S)^2$ as advertised~\cite{cremi}.

The basic result here is not surprising, $\omega_\star \simeq \dot \phi / f$ sets the energy scale of the oscillatory behavior which controls the largest values of $k_S/a(t)$ with which quanta can be produced.  As a result, using $k_L \gtrsim a(t)H$, we find that we need $\tfrac{k_S}{k_L} < \tfrac{\omega_\star}{H}$ if we want to violate the consistency condition. To make the regime in which the consistency relation is violated as large as possible, we must therefore increase $\omega_\star$ at fixed $H$. The value of $\dot\phi$ is fixed by the slow-roll potential, and we adjust $f$ to do this.  However, theories with an interaction as in Eq. (\ref{equ:resint0}) are a valid effective description for scales below the cutoff $\Lambda_U = 4\pi f$. The energy scales we study should all be below this cut-off~\cite{discrete}. In particular, we must satisfy $\omega_\star\leq \Lambda_U$, which implies that $f^2 \gtrsim \dot \phi$ and hence $\omega_\star < \dot \phi^{1/2}$. Once again, from the conditions $k_S/a(t) \leq k_\star/a(t) \simeq \omega_\star$ and $k_L \geq a (t) H$, we obtain 
\beq
\frac{k_S}{k_L} \lesssim \frac{\dot \phi^{1/2}}{H} \sim 10^{2}.
\eeq
Intuitively, from our discussion in Section \ref{sec:orig}, we understand that to produce a squeezed shape we are trying to create modes with energy $E \gg H$, by extracting it from the kinetic energy of $\phi(t)$.  By tuning the frequency in the cubic coupling, i.e. $\dot\delta (t) \simeq \cos(\omega_\star t)$, we can make this conversion possible for modes with higher momenta but we find that, for the validity of the effective description, we are ultimately limited by the kinetic energy density of the inflation, $\dot\phi^{1/2}$, which is the same scale that appears in the power spectrum.

\subsubsection{Dissipation}

Another example of non-adiabatic evolution arises when we allow $\phi$ to interact with additional (dissipative) degrees of freedom~\cite{trapped,unwind, warm}. In those scenarios the inflaton produces a steady burst of particles as it (slow) rolls during expansion. The dissipative effects allow for inflationary dynamics for a steep(er) potential. Interactions with these extra degrees of freedom will then excite the inflaton significantly, from the Bunch-Davies state to a semi-classical state. (For example, this occurs in warm inflation for $T \gg H$.) Even though in general dissipative effects require multi-fields, one can show that the consistency condition is obeyed for a vast class of models, provided these additional degrees of freedom do not contribute to curvature perturbations \cite{diseft2}.

The case of dissipation resembles that of excited initial states, although not as an initial condition but as a (recurrent) dynamical process, and one gets contributions to the three-point function that peak in the folded limit \cite{diseft}. However, since dissipative effects are more efficient when the `friction' coefficient $\gamma$ is much bigger than Hubble, the memory of the correlations is quickly lost and most contributions occur just before the modes freezeout. In practice this means there will not be a significant departure in the squeezed limit from what occurs in single-field inflation in a Bunch-Davies state. This was explicitly shown in \cite{diseft2}.

\subsection{Bogliubov initial states}\label{srexcited}
Let us now turn to excited initial states. As a cautionary remark, let us point out that for any model in which the inflationary phase lasted longer than the minimum number of e-folds required by the horizon problem, observationally we will not have access to the modes that were significantly excited at the beginning of inflation. So the discussion in this subsection is only relevant for models in which inflation lasted just long enough. 
Furthermore, for reasons we explain below let us restrict to a commonly studied but rather special class of excited states that can be obtained from the Bunch-Davies state by a Bogoliubov transformation. More general excited states will be discussed in~\cite{future}.

In these states, it is convenient to write the curvature perturbation as
\beq
\hat \zeta_\k(\tau) = (\alpha_\k \zeta_k(\tau) + \beta^*_{-\k} \zeta^*_{k}(\tau)) \hat b_\k + (\alpha^*_{-\k} \zeta^*_{k}(\tau) + \beta_{\k} \zeta_{k}(\tau)) \hat b_{-\k}^{\dagger} \,,
\eeq
where $b_{\k}^{\dagger}$ and $b_\k$ are creation and annihilation operators and $b_\k$ annihilates the state. The Bogoliubov coefficients satisfy $|\alpha_\k |^2 - |\beta_\k|^2 = 1$ and without loss of generality, we may take $\alpha_\k = \sqrt{1+\beta_\k^2} e^{i \theta_\k}$ with real $\beta_\k$. Up to slow-roll corrections, the mode function is given by 

\beq\label{eq:modefcn}
\zeta_k(\tau) = \frac{H^2}{\sqrt{2} \dot \phi} \frac{1}{k^{3/2}} (1+i k\tau) e^{-i k \tau} \,.
\eeq
As long as we limit ourselves to states with energies that are small compared to the kinetic energy of the inflaton, it is straightforward to compute correlation functions using these states~\cite{Boyanovsky:2006qi, Holman:2007na, Agullo:2010ws, Ganc:2011dy, Chialva:2011hc, Ganc:2012ae, Agullo:2012cs} and the power spectrum is 
\beq\label{equ:excitedpower}
k^3\langle \zeta_\k \zeta_{\k'}(0) \rangle'_\beta  \equiv \left(\Delta^\beta_\zeta\right)^2 = \frac{H^4}{2 \dot \phi^2} \Big(1 + 2 |\beta_\k|^2 + 2 \sqrt{|\beta_\k|^2 + |\beta_\k|^4} \cos(\theta_\k) \Big)  \,.
\eeq

The bispectrum is also readily calculated, though the final answer turns out to be somewhat lengthy~\cite{Holman:2007na, Agullo:2010ws, Ganc:2011dy, Chialva:2011hc, Ganc:2012ae, Agullo:2012cs}. For simplicity, we only present expressions for the squeezed limit for large and small $\beta_{\k_\star}$ as well as $\theta_\k\approx 0$.\footnote{A constant phase seems rather unphysical and does not arise in any microscopic model we know. However, at least for $\beta_{\k_\star}\gtrsim1$ this is required by scale invariance of the power spectrum, and we will allow for this possibility.} For large $\beta_{\k_\star}$ the squeezed limit of the three-point function becomes
\bea
\label{bibogo}
&&\langle \zeta_{\k_1} \zeta_{\k_2} \zeta_{\k_3} \rangle'_\beta |_{k_1 \ll k_2 , k_3< k_\star} \simeq 2 \epsilon  \left(\Delta^\beta_\zeta \right)^4 \frac{1-\cos(k_1\tau_0)}{k_1^4 k_2 k_3 } \qquad\hskip.55cm\text{for}\qquad \beta_{\k_\star}\gg1\,,\\
&&\langle \zeta_{\k_1} \zeta_{\k_2} \zeta_{\k_3} \rangle'_\beta |_{k_1 \ll k_2 , 
k_3< k_\star} \simeq 8 \epsilon \beta_{\k_\star}  \left(\Delta^\beta_\zeta \right)^4 \frac{1-\cos(k_1\tau_0)}{k_1^4 k_2 k_3 }\qquad\text{for}\qquad \beta_{\k_\star}\ll1 \,,
\eea
where $\tau_0$ is the conformal time at which the state was excited.
These expressions are useful to gain some intuition about the behavior of the bispectrum in the squeezed limit, but we will work with the full expression when we turn to forecasts for future experiments in section~\ref{sec:obs}.
We immediately see that we get an enhancement in the squeezed limit, provided $k_2 , k_3 < k_\star$. Notice also that the result is slow-roll suppressed. This can be understood from the fact that these states are a superposition of states containing only an even number of quanta. As a consequence, there are no three-point correlations built into the state and slow-roll interactions are required to generate them. This will not be true for general excited initial states making the Bogoliubov states somewhat pathological, if the goal is to discuss the possibility of a non-trivial three-point function. We postpone the study of general initial states to~\cite{future}.

\subsubsection{Bound on $k_S/k_L$ for Bogoliubov initial states}

Let us now derive the bound on $k_S/k_L$. From equation~\eqref{equ:excitedpower} we see that the requirement of scale invariance imposes $\beta_\k \sim \beta_{\k_\star}$ for (as well as a scale independent phase $\theta_{\k} \simeq \theta_{\k_\star}$) for all the modes we observe. Moreover, maintaining a finite energy density requires  $|\beta_\k|^2 \propto k^{-n}$ with $n>4$ when $k> k_{\star}$. \vskip 6pt

As explained before, the constraint on $k_\star$ comes from bounding the total energy density in the state
\beq
\label{00zeta}
\langle T^{\zeta}_{00} (\tau_0) \rangle \simeq \int^{k_{\star}} \frac{d^3 k}{(2\pi)^3} \frac{k}{a_0^4} |\beta_{\k_\star}|^2 \sim \tfrac{1}{8 \pi^2}  \frac{k_\star^4}{a_0^4} |\beta_{\k_\star}|^2 \,.
\eeq
Here $a_0$ corresponds to the value of the scale factor at the time $\tau_0$ when the modes are excited. If the energy density in this state is to leave the background evolution unaltered, it must be less than the kinetic energy of the inflaton. To see this explicitly, consider the equation for $\dot{H}$ in the presence of this energy density
\beq
\label{slowrollro}
-\Mp^2 \dot H \simeq \tfrac{1}{2} \dot \phi^2 + \tfrac{2}{3} \tfrac{1}{8 \pi^2} \frac{k_{\star}^4}{a_0^4}\left(\frac{a_0}{a}\right)^4 |\beta_{\k_\star}|^2 \,.
\eeq
So unless the second term on the RHS of this equation is subleading to the first, $\dot{H}$ is affected by the excited state, and to avoid this, we should impose
\beq
\label{kstarbound}
\frac{1}{6 \pi^2} \frac{k_{\star}^4}{a_0^4} |\beta_{\k_\star}|^2 \ll \dot\phi^2\,.
\eeq
Using the power spectrum in (\ref{equ:excitedpower}), the above inequality becomes
\beq
\label{kstarbound2}
\frac{k_{\star}^4}{(a_0 H)^4} \ll \frac{6\pi^2\dot \phi^2}{H^4 |\beta_{\k_\star}|^2}  \simeq 3\pi^2\left(\Delta^\beta_\zeta\right)^{-2} \frac{1+2|\beta_\k|^2+2\sqrt{|\beta_\k|^2 + |\beta_\k|^4} \cos(\theta_\k) }{|\beta_{\k_\star}|^2} \ .
\eeq
The simple lesson here is that the scales that control the power spectrum, namely $H^4$ and $\Mp^2 |\dot H| \simeq \dot \phi^2$, are the same scales that appear in the back-reaction constraint on $k_\star$ ($> k_S$).

Since $k_S \leq k_\star$ and $k_L \geq a_0 H$, we see that the constraint that the effect on the background evolution be negligible has enforced 
\bea  
&&\frac{k_S}{k_L} \lesssim 10^{2}\qquad\text{\hskip.25cm for}\qquad \beta_{\k_\star}\gg1\,,\\
&&\label{ratiok} \frac{k_S}{k_L} \lesssim \frac{10^{2}}{\beta_{\k_\star}^{1/2}}\qquad\text{for}\qquad \beta_{\k_\star}\ll1\,.
\eea  
Large occupation numbers are thus immediately ruled out because this bound would imply a step in the power spectrum, which is not observed. 

For small occupation numbers the bound is weakened and occupation numbers around $\beta_{\k_\star}\lesssim 10^{-1}$ certainly are compatible with the observed power spectrum. However, combining our bound~\eqref{ratiok} with equation~\eqref{bibogo}, we see that the amplitude of the three-point function in the squeezed limit decreases like $\beta_{\k_\star}^{1/2}$ as we decrease $\beta_{\k_\star}$. One might thus suspect that the amplitude will be too small to be detectable. We will be more quantitative about this and discuss the observability in more detail in section~\ref{sec:obs}, showing that this suspicion is correct.

\subsubsection{Generalization to small speed of sound}

One can easily generalize our previous argument to the case of small speed of sound models. The main difference is the computation in Eq. (\ref{00zeta}) for the energy density, which is now given by 
\beq
\langle T^{\zeta}_{00} \rangle \simeq \int^{k_{\star}} \frac{d^3 k}{(2\pi)^3} \frac{c_s k}{a^4} |\beta_{\k_\star}|^2 \simeq \tfrac{1}{8 \pi^2}  \frac{c_s k_{\star}^4}{a^4} |\beta_{\k_\star}|^2 \ .
\eeq
As in the case of slow-roll, this contribution must be negligible compared with $\Mp^2 \dot H$.  As a result, we find
\beq
\label{boundksold}
\tfrac{1}{8 \pi^2}  \frac{c_s k_{\star}^4}{a_0^4} |\beta_{\k_\star}|^2 <\frac32 \Mp^2 |\dot H| \ .
\eeq
The condition for freeze-out is modified due to the sound horizon, and therefore we now require $ k_L \gtrsim a_0 H/c_s$ for the long wavelength mode to be inside the horizon when the correlations are created. Using the power spectrum, and following the same steps as before, we obtain
\beq
\label{resultcs}
\frac{k_S^4}{k_L^4} < \frac{12\pi^2\Mp^2 |\dot H| c_s^3}{H^4|\beta_{\k_\star}|^2}  \simeq 3\pi^2\left(\Delta^\beta_\zeta\right)^{-2} c_s^2 \ .
\eeq
If we demand that there is no superluminal propagation, namely $c_s \leq 1$, then the bound on $k_S/k_L$ is at least as strong as before.

\section{Bounds on more general models: The EFT of inflation}\label{sec:eft}

In this section we explore excited initial states and non-adiabatic evolution from the perspective of the EFT of inflation, developed in \cite{Creminelli:2006xe, Cheung:2007st,Senatore:2010wk,diseft}. Here we will only cover the necessary steps to generalize the results from our previous section to include a wider class of models,  for instance those with a non-trivial dispersion relation. We will elaborate on the more formal aspects of the EFT formalism elsewhere \cite{future}. (See also \cite{Agarwal:2012mq}.)

The EFT of inflation is characterized by the spontaneous breaking of the time diffeomorphisms of the FRW background geometry.  As a result, one may write terms in the action that depend explicitly on time, provided the diffeomorphisms are realized non-linearly.  This can be achieved by introducing the St\"uckelberg field $\pi$, and writing the action as a function of $t + \pi$ such that it is now invariant under: $t \to t +\xi^0(x,t)$, $ \pi \to \pi -\xi^0(x,t)$. The action can then be written as 
\bea
S = \int d^4 x \sqrt{-g} \left[ \, \tfrac{1}{2} M_{\rm pl}^2 R - F(t+\pi) - c(t+\pi) \partial_\mu (t+\pi) \partial^\mu(t+\pi)+\cdots\right] \,.
\eea
The functions $F(t)$ and $c(t)$ are constrained by demanding that the background geometry is an FRW universe with metric
\beq
ds^2 = - dt^2 + a(t)^2 \left(\frac{dr^2}{1-\kappa r^2} + r^2 d \Omega^2\right) \ .
\eeq
It is straightforward to solve Einstein's equations. For situations when we only have one contribution to the energy budget of the universe one finds \cite{Creminelli:2006xe, Cheung:2007st} (setting $\kappa=0$)
\bea\label{Normalizations}
c(t) &=& - \Mp^2 \dot H \\
F(t) & =& \Mp^2\left(3 H^2(t) + \dot H(t)\right)\nonumber \ .
\eea
In the presence of an excited state, the energy density stored in the excitations, i.e. $\rho_\pi$, will also contribute to the energy budget and background equations, and may even fluctuate. In such a scenario one should distinguish between two type of perturbations, those associated with the `adiabatic mode', which we denote as $\tilde\pi$, and those which account for the perturbations of the clock that controls the end of inflation, which we will retain as $\pi$. For the former  \eqref{Normalizations} is still valid, whereas for the latter we have \cite{diseft,diseft2}
\bea\label{eq:rhopi}
c(t) &=& - \Mp^2 \dot H - \tfrac{2}{3} \rho_\pi(t), \\
F(t) & =& \Mp^2\left(3 H^2(t) + \dot H(t) - \tfrac{1}{3} \rho_\pi(t)\right)\,.
\eea
In this paper, we work in the regime $\rho_\pi\ll\Mp^2 |\dot H|$ so that the two are essentially the same $\tilde\pi\approx\pi$. The regime $\rho_\pi\simeq \Mp^2 |\dot H|$ for which the distinction is important will be studied in~\cite{future}.  

\subsection{Non-adiabatic evolution}

In section \ref{nonadia} we discussed resonant non-Gaussianity which, in the language of the EFT of Inflation, arises from the time dependence of $H(t) \sim H_0(t) + H_{\rm osc}(t) \cos(\omega_\star (t+\pi))$.  The key feature that leads to a non-trivial squeezed limit was the rapid time dependence from $\omega_\star \gg H$.  We can now generalize this analysis to any form of non-adiabatic evolution by considering generic time-dependent couplings.  We will focus here on models that lead to non-trivial squeezed limits and study the constraint on the number of modes involved from requiring the EFT is weakly coupled.

Consider adding a contribution to the Lagrangian for $\pi$ of the form
\beq\label{equ:resint}
\delta {\cal L}_{\rm int} \supset M^4 f(t+\pi)~g(\partial_\mu (t+\pi), \partial^2 (t+\pi), \cdots) \ ,
\eeq
where $\dot f(t) \sim \omega_\star f(t)$, $g(\partial_\mu (t+\pi), \partial^2 (t+\pi), \cdots)$ is polynomial in its arguments and $M$ is some mass scale required by dimensional analysis. In the spirit of generality and due to the natural features of the models, here we will deal with an energy scale $\omega_\star$ rather than a momentum scale $k_\star$, as we did in section \ref{nonadia}, to allow for generic forms of the dispersion relation, i.e. $\omega_\star = c_p (\omega_\star) k_\star$.\vskip 6pt

We will be interested in the case where the correlation functions are nearly scale invariant, and therefore impose an approximate (discrete) shift symmetry: $f(t + c) \sim f(t)$ \cite{discrete}.  As a result, we expect that at a generic instant of time: \beq \tfrac{\partial^n}{\partial t^n} f(t) \sim \omega_\star^n f(t).\eeq  These criteria imply that the scenario we have in mind is effectively a generalization of resonant non-Gaussianity \cite{Flauger:2010ja} (some such generalizations were discussed in \cite{discrete, Behbahani:2012be}).

We will take the kinetic term for $\pi$ to be of the form ${\cal L}_{\rm kintetic} = \Lambda^4 \dot \pi^2$, such that the canonically normalized field is $\pi_c = \Lambda^2 \pi$. (In slow-roll inflation, one has $\Lambda^4 \simeq \dot \phi^2$.) Expanding Eq. (\ref{equ:resint}) in powers of $\pi$ then we find
\beq
\label{Lintad}
\delta {\cal L}_{\rm INT} \supset M^4 \omega_\star^n \pi^n f(t)~g(\partial_\mu (t+\pi), \partial^2 (t+\pi), \cdots) = M^4 \Big( \frac{ \pi_c \omega_\star}{\Lambda^2}\Big)^n \, g(\partial_\mu (t+\pi), \partial^2 (t+\pi), \cdots) 
\eeq
In the limit of large $n$, the details of $g$ are irrelevant for determining the strong coupling scale $\Lambda_{\rm strong} \sim \Lambda^2/ \omega_\star$.  As in section \ref{nonadia}, the contributions to the squeezed limit arise at energies of order $\omega_\star$. 

\subsubsection{Generalized bound on $k_S/k_L$}

In order for the expansion in \eqref{Lintad} to be under control, we must require 
\beq \label{equ:constraint}
\frac{\pi_c |_{\omega \sim \omega_\star} \omega_\star}{\Lambda^2} \ll 1 \ .
\eeq
If this inequality is violated, the effective theory is strongly coupled at the scale $\omega_\star$ and therefore the bispectrum is incalculable. Therefore, to generalize our results from (\ref{nonadia}) we need to place an upper bound on $\Lambda$.  If there is no such bound, one could make $\omega_\star$ arbitrarily large without violating (\ref{equ:constraint}).  Following \cite{ArkaniHamed:2007ky, Baumann:2011ws}, one can show that the null energy condition implies  
\beq
\Mp^2 | \dot H| \geq \Lambda^4 \dot \pi  |_{\omega \sim \omega_\star} =  \Lambda^2  \dot \pi_c  |_{\omega \sim \omega_\star} > \omega_\star^2 \pi_c^2 |_{\omega \sim \omega_\star}  \ ,
\eeq
where in the last step we used (\ref{equ:constraint}) and $\dot \pi_c = \omega \pi_c$.  Our theory is nearly Gaussian with a kinetic term $\dot \pi_c^2$, therefore $\pi_c \sim k^{\tfrac{3}{2}} \omega^{-1/2} =  c_p^{-\tfrac{3}{2}} \omega$ at high energies, which leads to 
\beq
\Mp^2 | \dot H| > \omega_\star^4 c_p^{-3} (\omega_\star) \ .
\eeq
Therefore, correlations between short and long momenta can only occur in the range
\beq
\frac{k_S}{k_L} \lesssim \frac{c_p(\omega \simeq H)}{c_p(\omega \simeq \omega_\star)} \frac{\omega_\star}{H} < \frac{(\Mp^2 |\dot H|)^{1/4}}{H} \frac{c_p(H)}{c_p^{1/4}(\omega_\star)} \lesssim \Delta_\zeta^{-1/2}\frac{c^{1/4}_p(H)}{c_p^{1/4}(\omega_\star)} \lesssim 10^{2} ,
\eeq
where we used (see appendix \ref{appB}) \beq \Mp^2 |\dot H| / H^4 < \tfrac{1}{4} \Delta_{\zeta}^{-2} c^{-3}_p(\omega \simeq H) \ . \eeq  
In the final step, we have ignored the ratio of phase velocities, which is justified by the small fractional power and that the phase velocities grow with energy for simple dispersion relations.
\subsection{Bogoliubov initial states}\label{sec:excited1}

Let us now generalize our bounds derived in section~\ref{srexcited}. As before, we restrict ourselves to the case in which the energy density of the excited state does not affect the background evolution. From equation~\eqref{eq:rhopi}, we see that this implies 
\beq
\label{rhopibound}
\rho_\pi(t) \ll \tfrac{3}{2} \Mp^2 |\dot H|\,.
\eeq  

\subsubsection{Generalized bound on $k_S/k_L$}

Starting from (\ref{rhopibound}), we can now follow the steps in \ref{srexcited} to write this inequality as
\beq\label{equ:energy}
\rho_\pi (\tau_0) =\langle T^{(\pi)}_{00}(\tau_0) \rangle \simeq \frac{1}{a(\tau_0)^3} \int^{k_\star} \frac{d^3 k}{(2\pi)^3} \omega(\tfrac{k}{a(\tau_0)}) |\beta_\k|^2 \simeq \tfrac{1}{8 \pi^2} \frac{ k_\star^3 \omega(\tfrac{k_\star}{a(\tau_0)})}{a(\tau_0)^3} |\beta_{\k_\star}|^2  < \tfrac{3}{2} \Mp^2 |\dot H| \ .
\eeq
The main difference with Eq. (\ref{00zeta}) is the scaling, because here we did not assume a dispersion relation of the form $\omega \simeq k$. We can now show how our bound holds more generally.  Specifically, we can always write (\ref{equ:energy}) as 
\beq\label{equ:bound}
\frac{ \omega^4(\tfrac{k_\star}{a(\tau_0)})}{H^4 c_p^3(k_{\star}, \tau_0)} |\beta_{\k_\star}|^2  < 12 \pi^2 \frac{\Mp^2 |\dot H|}{H^4} \ ,
\eeq
where $c_p(k,\tau) = \tfrac{\omega(k/a)}{(k/a)}$ is the phase velocity.  Moreover (see Appendix \ref{appB})
\beq
\label{ineqcp}
\frac{\Mp^2 |\dot H|}{H^4} < \tfrac{1}{4}\left(\Delta^\beta_{\zeta}\right)^{-2} c_p^{-3}(\omega \sim H) \Big[1+2|\beta_\k|^2+2\sqrt{|\beta_\k|^2 + |\beta_\k|^4} \cos(\theta_\k) \Big] \,.
\eeq
This inequality implies that the right hand side of (\ref{equ:bound}) is bounded by the power spectrum. Using once again $k_S \lesssim k_\star$ we obtain, for $|\beta_{\k_\star} |^2 > 1$ (and barring tuned cancellations in the power spectrum),
\beq
\frac{ \omega^4(\tfrac{k_S}{a(\tau_0)})}{H^4 c_p^3(k_S, \tau_0)} < 6 \pi^2 \left(\Delta_\zeta^\beta\right)^{-2} c^{-3}_p(k_L,\tau_0) \ .
\eeq
We can then use this inequality to generalize our bound for the ratio of scales: 
\beq
\label{resultcp}
\frac{k_S}{k_L} = \frac{c_p(k_L)}{c_p(k_S)} \frac{\omega(k_S)}{H} < \Big( \frac{c_p(k_L)}{c_p(k_S)}\Big)^{1/4} \left(\Delta^\beta_{\zeta}\right)^{-1/2} \lesssim 10^2 \ .
\eeq 
Hence, even in the presence of a non-trivial dispersion relation, the bound remains essentially unchanged. 

The reader may have already noticed that the bound in (\ref{resultcs}) was somewhat stronger than the one above. In other words, (\ref{resultcp}) does not reproduce (\ref{resultcs}) when $c_p=c_s$. The reason is that here we used inequality in (\ref{ineqcp}), which is valid regardless of the form of the dispersion relation; whereas in specific examples one can derive the form of the power spectrum and $\Delta^\beta_{\zeta}$ precisely, which in the case of small speed of sound models turns out to be a factor of $c_s^2$ from saturating this bound. In any case, we recover the condition $\frac{k_S}{k_L} \lesssim 10^{2}$ for a broader class of models, as anticipated.

\section{Prospects for observations }\label{sec:obs}

We now discuss the observational constraints on the models discussed in previous sections.

\subsection{Non-adiabatic evolution}

The signatures of resonance models have been discussed in the literature, so we will simply summarize the relevant results.

The strong-time dependence that allows us to violate the consistency condition also tends to produce large violations of scale invariance in the power spectrum.  This is not surprising, as scale invariant correlation functions arise when each mode follows the same history.  The time dependence of the couplings means that modes that cross the horizon at different times see different values of the couplings. It is this violation of scale invariance in the power spectrum that is the dominant signature of these models \cite{Flauger:2010ja,discrete} rather than the violation of the consistency condition for some intermediate regime of scales. This is not special to the specific realization of the model but is a generic feature of resonant models \cite{discrete}.  The reason is fairly intuitive. Once the approximate shift symmetry is broken, generically nothing forbids a contribution to the power spectrum at the same order as in the higher correlation functions.  Then, since the spectrum is approximately Gaussian, the power spectrum will dominate the signal-to-noise ratio.

One can try to avoid the above conclusion by introducing additional symmetries that forbid the large contributions to the power spectrum.  Models of this type in which the bispectrum did contain most of the signal-to-noise were constructed in \cite{Behbahani:2012be}. However, the models found there did not lead to a non-trival squeezed limit. The reasons are similar to what one finds for excited states with small $c_s$, namely that the large interaction terms that increase the relative significance of the bispectrum are accompanied by derivatives that suppress the squeezed limit, even inside the horizon.

For the case of dissipation, as we mentioned earlier, the scaling of the three-point function is not modified in the squeezed limit \cite{diseft2}, however, one finds a peak at the folded configuration \cite{diseft}. This is connected with the fact that particles are being produced during inflation \cite{warm,trapped,unwind}.  This enhancement for the folded shape of the bispectrum occurs for {\it any} excited state, motivating the exploration of this (and other) shapes in the data.\\

In summary, we are not aware of any model that produces a non-trivial squeezed limit and at the same time does not lead to a large signal in the power spectrum. 

\subsection{Bogoliubov initial states}

The excited initial states we have discussed were related to the Bunch-Davies state by a Bogoliubov transformation. As we mentioned earlier, the regime $\beta_{\k}\gg1$ is ruled out by observations of the power spectrum. To see this recall that the power spectrum takes the form
\beq
k^3\langle \zeta_\k \zeta_{\k'}(0) \rangle' =  \frac{H^4}{2\dot\phi^2} \Big(1 + 2 |\beta_\k|^2 + 2 \sqrt{|\beta_\k|^2 + |\beta_\k|^4} \cos(\theta_\k) \Big) \,.
\eeq
As we explained, negligible backreaction on the background evolution enforces \[k_S / k_L \lesssim \Delta_\zeta^{-1/2} \sim 10^{-2},\] and we can only significantly excite modes over two orders of magnitude. For $k > k_\star$ the Bogoliubov coefficients must decrease at least like $\beta_\k \propto k^{-n}$ with $n>4$. For generic values of  $\theta_\k$, this implies that the power spectrum scales like $P_\zeta \propto k^{-3-n}$ for $k> k_\star$, leading to a step in the power spectrum of primordial scalar perturbations. Observations of the cosmic microwave background alone by WMAP~\cite{wmap}, SPT~\cite{spt} and ACT~\cite{act}, have reliably measured the angular power spectrum of temperature anisotropies over a range that is large enough to rule this out. We thus conclude that $\beta_{\k_\star} \gtrsim 1$ is incompatible with existing data.

There appears to be a loophole in the above analysis for the case of Bogoliubov states when $\cos(\theta_\k) = -1$ and $\beta_\k \gg 1$. The power spectrum becomes independent of $\beta_\k$ up to $\beta_\k^{-2}$ corrections.  However, to make use of this to avoid our conclusions, one would still need $\beta_\k \gg 1$ for all observable modes. For $k>k_\star$ we have $|\beta_\k|^2 \propto k^{-n}$ with $n>4$. Therefore, to accomodate one more decade of highly excited modes to be compatible with observations, one needs $\beta_{\k_\star}  \gg 10^2$.  However, this large occupation number is only compatible with the bound on the energy density of the excited modes, which takes the form $|\beta_{\k_\star}|^2 \tfrac{k_\star^4}{a_0^4} \ll \dot\phi^2$ (see (\ref{kstarbound})), if $k_\star/k_L\lesssim 10$, in contradiction with the observation of the power spectrum. One may wonder whether other types of cancellations could occur for more complicated excited states. Again, this cancellation is most likely to occur for large occupation numbers, in which case our previous reasoning applies in the same manner. (See \cite{future} for more details.)

For the case of $\beta_{\k_\star} \ll 1$, the constraints are weaker for two reasons.  First, the contribution to the power spectrum is sub-leading relative to the scale invariant contribution, as it is nearly in the vacuum state.  Second, the lower occupation numbers weaken the constraint on $k_S$, allowing for the possibility of all the observed modes in an excited state with low occupation number $\beta_\k \sim \beta_{\k_\star} \ll 1$. Nevertheless, the question remains if states with $\beta_{\k_\star} \ll 1$ could produce a measurable effect in the squeezed limit.  We now discuss two (in principle) possible routes for observation.

\subsubsection{Observing the squeezed limit in the CMB}

For the CMB, we can gain fairly reliable intuition for the signal-to-noise ration $S/N$ by working with the correlation functions directly.  In appendix~\ref{append:sn} we derive the signal-to-noise ratio for a direct measurement of the bispectrum,
\beq
\label{snexcited}\Big(\frac{S}{N} \Big) \sim 2\epsilon \beta_{\k_\star}\frac{k_S}{k_L} \times  \Delta_\zeta \sqrt{{\cal N}_\text{pix}} \,,
\eeq
where ${\cal N}_\text{pix}$ is the number of modes measured. A direct measurement of the bispectrum of primordial curvature perturbations is of course unrealistic, but  incorporating effects of transfer functions, sky cuts, etc. necessarily decreases the signal-to-noise ratio so that this surely provides an upper bound. On the other hand, from the bound in equation~\eqref{ratiok} we notice that, for small occupation number, at best $k_S/k_L\sim 10^2\beta_{\k_\star}^{-1/2}$ so that we do not gain any signal by decreasing the occupation numbers to increase $k_S/k_L$.  Therefore, the largest signal is expected\footnote{We take this value as an upper limit consistent with the approximate scale invariance of the power spectrum.} for $\beta_{\k_\star}^2  \lesssim 10^{-1}$ and $k_L / k_S \sim 10^{-2}$. Similarly, slow-roll inflation requires $\epsilon \lesssim 10^{-2}$. Hence, using the number of observable modes in the CMB, ${\cal N}_\text{pix} \sim \ell_{\rm max}^2 \sim (2000)^2$ and $\Delta_\zeta = 2.2 \times 10^{-4}$, one finds
\beq
\Big(\frac{S}{N} \Big) \lesssim \frac{1}{3} \ .
\eeq
So even with the most optimistic choices of parameters, we see that no signal is expected with Planck \cite{planck}.

We might hope to increase the signal at small $\beta_{\k_\star}$ by introducing interactions with large coefficients.  We study this possibility in Appendix \ref{append:sn} for small $c_s$ models.  In the absence of an excited state, these models have an enhanced $f^{\rm equilateral}_{\rm NL} \propto c_s^{-2}$ and can therefore lead to a detectable signature in equilateral configurations ($k_1 \sim k_2 \sim k_3$).  However, even in the presence of an excited state, there is no significant improvement for the signal-to-noise in the squeezed limit. Therefore, even allowing for stronger cubic interactions, these models cannot produce measurable deviations from the consistency condition in the CMB.

\subsubsection{Observing the squeezed limit in LSS}

Large scale structure surveys \cite{lss} have emerged as a powerful probe of squeezed limits of non-Gaussian correlation functions~\cite{Dalal:2007cu}.  However, unlike the CMB, it is less clear what range of momenta contribute to the signal-to-noise for a given experiment and shape.  For this reason, we will explore a forecast for a Euclid-like LSS survey~\cite{euclid}, rather than using the simplified analysis we followed for the CMB. A detailed forcast is certainly beyond the scope of this paper and not even currently possible because the specifications of the mission are still evolving. However, using the specifications from the Red Book~\cite{Laureijs:2011mu} as a guide for a simple forcast should allow us to draw robust conclusions especially because incorporating details would only decrease our idealistic signal-to-noise estimates. 

We expect the largest signal-to-noise ratio for a detection of the Bogoliubov initial states with low occupation numbers to arise from the the galaxy bispectrum. However, since we do not expect significant changes to our conclusions, we will focus instead on the measurement of the 3D power spectrum 
\beq
{\cal P}_g(k) = b_g^2 {\cal P}_m(k)\,.
\eeq
Here ${\cal P}_m(k)$ is the underlying matter power spectrum and $b_g$ is the galaxy bias. 

\noindent The power spectrum is usually measured from the spectroscopic part of the survey. Euclid is expected to use a slitless spectrometer with a wavelength range \mbox{$1100\,\text{nm}<\lambda<2000\,\text{nm}$}. The redshift measurement will rely on emission lines of galaxies, predominantly H$\alpha$, so that this wavelength range translates to a redshift range for the observed galaxies of $0.7\lesssim z\lesssim 2.1$. With a limiting flux and success rate of the spectrometer as given in~\cite{Laureijs:2011mu}, one expects a sample of $52.5\times 10^6$ galaxies over the survey area of the ``required'' $15,000$ square degrees. We will be optimistic and use the ``goal'' of $20,000$ square degrees rather than the ``required'' $15,000$ square degrees for our forecast. 

\begin{figure}[t!] %  figure placement: here, top, bottom, or page
 \begin{center}
 \includegraphics[width=5.5in]{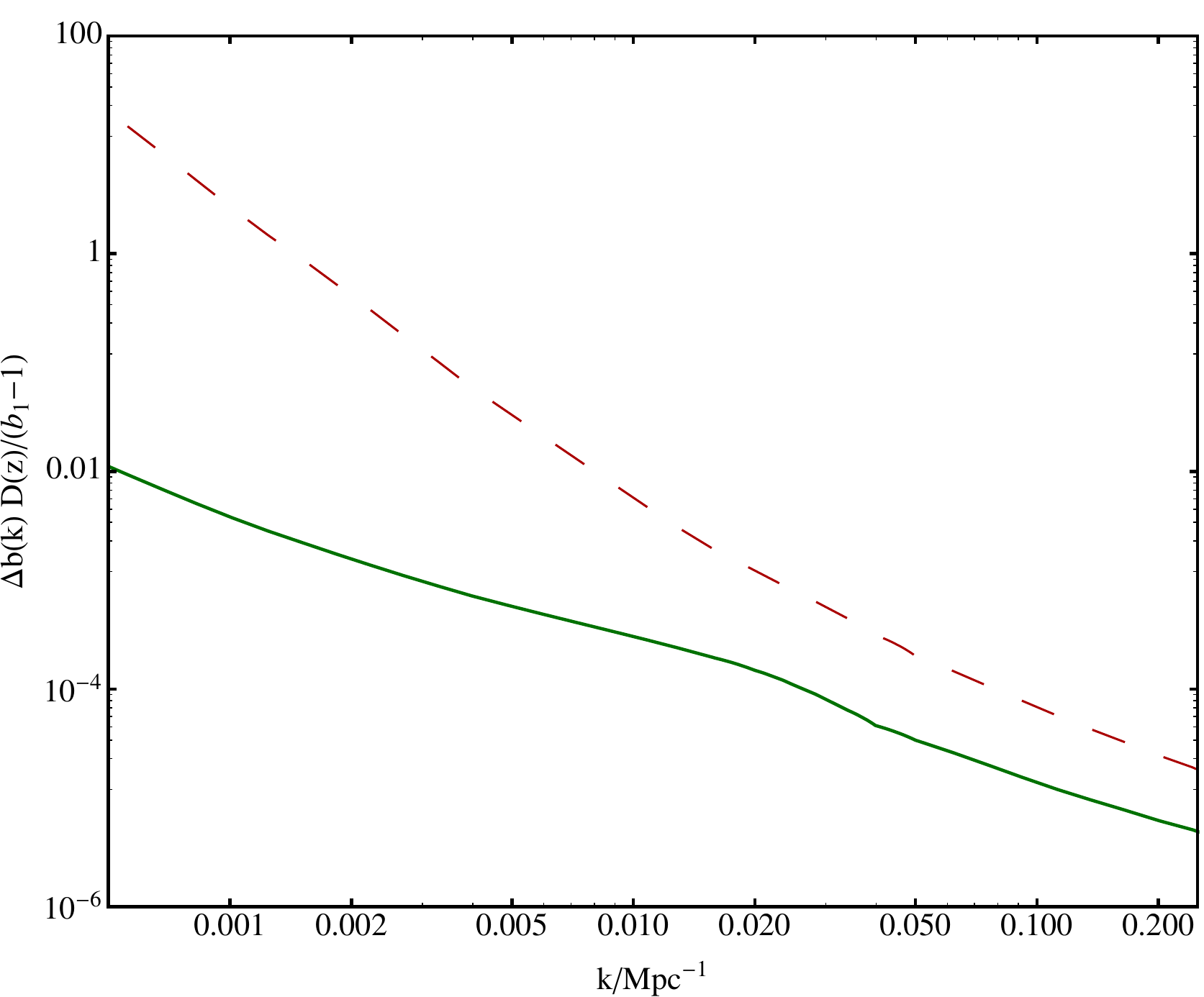}
 \caption{The green and red curves show the scale dependent halo bias as a function of $k$ with and without the bound~\eqref{ratiok} enforced, respectively. We have chosen $M=10^{12} h^{-1}\, M_\text{sol}$, $\beta_{\k_\star}=0.3$, $k_\star=1\,Mpc^{-1}$, and slow-roll parameters $\epsilon=\eta=10^{-2}$.}
 \label{fig:dbofk}
 \end{center}
\end{figure}

Primordial non-Gaussianity has several effects on the galaxy power spectrum. The dominant effect is to introduce a scale-dependence into the bias 
\beq
b_g(k,f_{NL}) = b_g(f_{NL}=0)+\delta b_g(f_{NL})+\Delta b_g(k,f_{NL})\,.
\eeq
According to the halo model, the galaxy bias is related to the underlying halo-bias by
\beq\label{eq:effbias}
b_g(k,f_{NL})= \int\limits_{M_\text{min}}^\infty dM\, b_h(k,f_{NL},M) \frac{n(M) N_g(M)}{n_g}\quad\text{with}\quad n_g=\int\limits_{M_\text{min}}^\infty dM\, n(M) N_g(M)\,,
\eeq
%\beq
%b_g(k,f_{NL})\propto \int dM\, b_h(k,f_{NL},M) n(M) N_g(M)\,,
%\eeq
where $N_g(M)$ is the number of galaxies expected in a halo with mass $M$, and $n(M)$ is the number density of halos with mass between $M$ and $M+dM$.  The main lesson we take from these equations is that the scale-dependent contribution of the galaxy bias inherits its properties from the scale-dependent halo bias, which can be calculated as in~\cite{Schmidt:2010gw,Desjacques:2011mq,Desjacques:2011jb,Ferraro:2012bd,Baumann:2012bc}
\beq
\Delta b_h(k,f_{NL},M) =  \frac{1}{\mathcal{M}_M(k)}\left[\frac{(b_1-1)\delta_c}{D(z)}\mathcal{F}(k,M)+ \frac{d\mathcal{F}(k, M)}{d\log \sigma_M} \right] \,,
\eeq
where $b_1$ is the effective (Eulerian) Gaussian halo bias, $\delta_c$ is the collapse threshold, and $\sigma_M$ is the matter power spectrum smoothed on the scale $R=\left(2M G/\Omega_m H_0^2\right)^{1/3} $.  Furthermore, $D(z)$ is the linear growth function normalized to unity at redshift zero, $\mathcal{M}_M(k) = k^2 T(k) W(kR)$ where $T(k)$ is the transfer function in the conventions used by CAMB, and $W(x)=3(\sin x-x \cos x)/x^3$ is the Fourier transform of a top-hat function. Finally,
\beq
\mathcal{F}(k, M) = \frac{1}{2\sigma_M^2 \langle|\zeta_\k|^2\rangle'} \int\frac{d^3 p}{(2\pi)^3}  \mathcal{M}_M(p) \mathcal{M}_M(|\k +\p|) \langle \zeta_\p \zeta_\k \zeta_{-\p-\k} \rangle'\,.
\eeq
The resulting halo bias generated by the excited states discussed in Section \ref{srexcited} was computed in \cite{Ganc:2012ae, Agullo:2012cs}. We show it in figure~\ref{fig:dbofk} for a halo mass of $10^{12}h^{-1}M_\text{sol}$, where the green and red lines represent the halo bias with and without our bound~\eqref{ratiok} enforced, respectively. \\

Let us now estimate what we can observe from a measurement of the power spectrum in our Euclid-like survey. We will assume a Gaussian likelihood function in the various momentum and redshift bins and calculate
\beq\label{eq:dchi21}
\Delta\chi^2=\sum\limits_{k,i}\frac{\overline{\cal P}_g^2(k,z_i)}{\sigma^2(k,z_i)}\frac{({\cal P}_g(k,z_i)-\overline{\cal P}_g(k,z_i))^2}{\overline{\cal P}_g^2(k,z_i)}=\sum\limits_{i}\int k^2dk \frac{V_\text{eff}(k,z_i)}{(2\pi)^2}\frac{\Delta {\cal P}_g(k,z_i)^2}{\overline{\cal P}_g^2(k,z_i)}\,,
\eeq 
where $\overline{\cal P}_g$ denotes a fiducial galaxy power spectrum. For the last equal sign, we have traded the sum over momenta for an integral, and we have used that the cosmic variance for the power spectrum measurement, up to a factor of 2, is just given by the effective number of modes that enter into the measurement (see {\it e.g.}~\cite{Feldman:1993ky})
\beq
\frac{\sigma^2(k,z_i)}{\overline{P}^2(k,z_i)}=\frac{2(2\pi)^3}{4\pi k^2\Delta k V_\text{eff}(k,z_i)}\,.
\eeq 
Here $V_\text{eff}(k,z_i)$ is the effective volume. Especially for a flux limited survey like Euclid it may be significantly smaller than the geometric volume \cite{Orsi:2009mj}.\footnote{ 
As shown in~\cite{Orsi:2009mj}, the effective volume for different semi-analytic models of galaxy formation for $H\alpha$ selection may be as small as a few per cent of the geometric volume for realistic Euclid flux limits and success rates.} 
We will nevertheless use the geometric volume for our estimates, which will surely lead to an optimistic signal-to-noise estimate. We fix all cosmological parameters characterizing the background to their best-fit values~\cite{wmap}, and use a fiducial power spectrum derived assuming a Bunch-Davies state. Keeping only the effect of the excited state on the scale-dependent bias, equation~\eqref{eq:dchi21} then simplifies to
\beq\label{eq:dchi2}
\Delta\chi^2=\sum\limits_{i}\frac{V(z_i)}{(2\pi)^2}\int_{k_\text{min,i}}^{k_\text{max,i}} k^2dk \left[\left(1+\frac{\Delta b_g(k,z_i)}{b_g(z_i)}\right)^2-1\right]^2\,.
\eeq 
\begin{figure}[t!] %  figure placement: here, top, bottom, or page
 \begin{center}
 \includegraphics[width=5.5in]{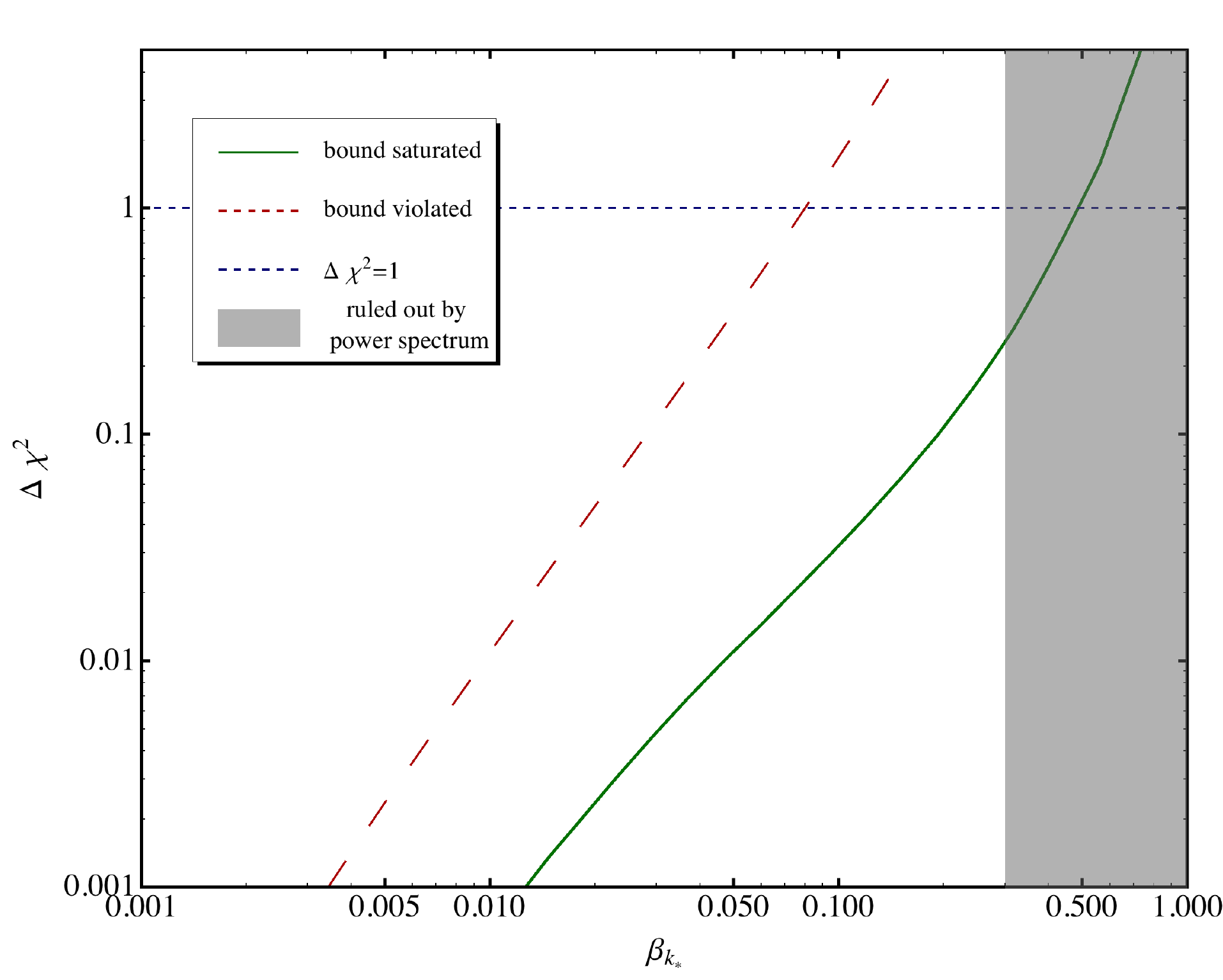}
 \caption{This plot shows $\Delta\chi^2$ as defined in~\eqref{eq:dchi2} as a function of $\beta_{\k_\star}$ with the phase $\theta_\k$ and $k_\star$ chosen to maximize the signal. The red and green curves show the results without and with the bound~\eqref{ratiok}, respectively. The shaded region indicates the region that is ruled out by the power spectrum when~\eqref{ratiok} is enforced.}
 \label{fig:dchi2ofb}
 \end{center}
\end{figure}
In a more conservative treatment, one should treat the Gaussian bias as nuisance parameter. Instead, we model it using a Gaussian halo bias appropriate for the Sheth-Tormen halo mass function~\cite{Sheth:1999mn,Sheth:1999su,Sheth:2001dp} and compute the effective galaxy bias from~\eqref{eq:effbias}. We use the Sheth-Tormen halo mass function for $n(M)$, and use a simple fitting form for the expected number of galaxies in a halo with mass $M$~\cite{Sheth:2000ii} 
\beq
N_g(M)=N_{g,\text{Blue}}\left(\frac{M}{M_\text{Blue}}\right)^{\alpha_B}\,,
\eeq
with $N_{g,\text{Blue}}=0.7$, $M_0=4\times 10^{12} M_\text{sol}h^{-1}$, and $\alpha_B=0$ or $\alpha_B=0.8$ for masses below or above $M_\text{blue}$, and we set $M_\text{min}=10^{11}M_\text{sol}h^{-1}$ in equation~\eqref{eq:effbias}.  This roughly reproduces the results found for the effective Gaussian bias of an $H\alpha$ sample typical for Euclid found for semi-analytic galaxy formation models in~\cite{Orsi:2009mj}.

We split the survey into 12 redshift bins and set the limits of integration in~\eqref{eq:dchi2} to $k_\text{min,i}=2\pi/V(z_i)^{1/3}$ as well as \mbox{$k_\text{max,i}=0.1 h\,Mpc^{-1}$}. We excite modes up to $k_\star$ over a range as large as allowed by our bound~\eqref{ratiok} choosing $k_\star$ as well as the phase $\theta_\k$ to maximize~\eqref{eq:dchi2}, and set $\epsilon=\eta=10^{-2}$. The result is shown in figure~\ref{fig:dchi2ofb}. 

We see that even though we have made a number of optimistic assumptions, the signal is undetectably small at least in the regime that is not already ruled out by power spectrum constraints.

\section{Discussion}\label{sec:disc}

In this paper we have studied single-field inflation with excited states arising either from non-adiabatic evolution or as remnants from a pre-inflationary epoch. We have shown that all existing models in which the state is continuously excited during inflation can only violate the consistency condition for a finite range of momenta bounded by $k_S/ k_L \lesssim 10^{2}$. This implies that the violation of the consistency condition is unobservable in these scenarios. 
For resonant models and generalizations thereof the violation of the consistency condition comes with periodic features in the power spectrum, which provide the dominant signature in these models. 
Strongly dissipative systems give rise to a scale-invariant power spectrum, but the consistency condition applies roughly in the same fashion as in single-field models with Bunch-Davies states, which in practice means that the squeezed limit is not useful to distinguish between the two and other corners of the bispectrum need to be explored, in particular folded configurations \cite{diseft,diseft2}. 
For Bogoliubov initial states with significant occupation numbers, but an energy density that is small enough so that the slow-roll conditions are satisfied (in particular $\delta=\tfrac{\ddot H}{2\dot H H} \ll 1$), we have also shown that the consistency condition can only be violated for a finite range of momenta bounded by $k_S/ k_L \lesssim 10^{2}$. Thus in this case the violation of the consistency condition comes with a step in the power spectrum in contradiction with observations. 
For Bogoliubov initial states with small occupation numbers, the constraint is weaker. However, the three-point function is generated by gravitational interactions for these states. As a consequence, as we show in the form of signal-to-noise estimates, the violation of the consistency condition is too small to be observed in Planck or Euclid.\\

What prevented us from extending our conclusions to models in which the energy density in the excited state, $\rho_\pi$, is larger than the kinetic energy in the inflaton (so that the condition $\delta \ll 1$ is 
violated) is the potential breakdown of a controlled theoretical framework. In this case the bound $k_S/ k_L \lesssim 10^{2}$ generically no longer applies, and a weaker bound on $k_S/k_L$ might lead to observable consequences in the bispectrum without jeopardizing the scale invariance of the power spectrum over the observed range of scales. This regime therefore deserves further scrutiny, and one has to carefully study all sources of density perturbations and relevant scales in the $n$-point correlation functions, while ensuring that perturbation theory remains under control. 

Even though we may consider models when $\rho_\pi$ exceeds the kinetic energy in the inflaton, we should always impose the so-called {\it back-reaction} constraint, $\rho_\pi < M_p^2H^2$, since its violation would spoil inflation at least for some period of time. (In fact, using the null energy condition we can show more generally: $\tfrac{2}{3}\rho_\pi \lesssim M_p^2|\dot H|$.\footnote{This simply follows in the EFT language from \eqref{eq:rhopi} and imposing $c(t)>0$, or \eqref{slowrollro} and the fact that $\dot\phi^2>0$.}) What is no longer straightforward in the regime $\rho_\pi \gg \dot\phi^2$ is how to relate this bound to the power spectrum. However, let us make two important observations about this regime. Firstly, if all the observed modes exit after $\rho_\pi$ has redshifted away (dropped below $\dot\phi^2$), our conclusions remain unaltered. Secondly, since during the non slow-roll phase $\epsilon \simeq a^{-4}$, we have $k_1/k_0 \simeq \left(\epsilon_0/\epsilon_1 \right)^{1/4}$ for modes with co-moving momenta $k_{0(1)}= a_{0(1)} H$.  This means that, for each decade of modes that exit during this phase the value of $\epsilon$ in the subsequent slow-roll epoch decreases by a factor of $10^4$. Recall that at most two decades of excited modes with large occupation number can exit during the slow roll phase. Therefore, in order to accommodate the observed $10^3$ number of modes we would need $\epsilon_\phi \lesssim 10^{-4}$ (taking $\epsilon_0 \lesssim 1$). This weakens the size of the (gravitational) interactions that produce the bispectrum for Bogoliubov states. Conversely, large field models with $|\beta_k| > 1$ cannot accommodate all three decades of (scale invariant) observed modes. We will discuss this and other aspects in detail in a forthcoming publication \cite{future}. \\

\noindent{\bf NOTE:} While this work was being completed we were informed by A. Aravind, D. Lorshbough, and S. Paban of their related work~\cite{sonia}. 

\acknowledgments
We thank Matias Zaldarriaga for helpful discussions. RF would also like to thank Eiichiro Komatsu for organizing the conference ``Critial Tests of Inflation Using Non-Gaussianity'' where preliminary results of this work were first presented~\cite{conf}. RF~is supported by NSF under grant NSF-PHY-0855425 and NSF-PHY-0645435. DG~is supported by the Stanford ITP and by the DOE under grant DE-AC02-76SF00515. RAP is supported by NSF under grant AST-0807444 and DOE grant DE-FG02-90ER40542.

\addtocontents{toc}{\protect\setcounter{tocdepth}{1}}
\appendix

\section{A useful inequality}\label{appB}

In the main text, we were able to bound the ratio of scales $k_S / k_L$ in terms of $\Mp^2 |\dot H|$.  In order to achieve an upper bound we must then find an upper bound on $\Mp^2 |\dot H|$ in terms of $H$ and fixed numerical coefficients.  We will show this is possible in this appendix.

Let us make the simplifying assumption that the kinetic term for $\pi$ is of the form $\Lambda^4 \dot \pi^2$.  One can show that our result generalizes to arbitrary kinetic terms, but this will be unnecessary for our purposes.  In the EFT of inflation this kinetic term arises from two separate operators in the Lagrangian
\beq\label{equ:kinetica}
{\cal L}_{\rm kinetic} =  \Mp^2 \dot H \partial_\mu (t+\pi) \partial^\mu (t + \pi) + \frac{M_2^4}{2} [\partial_\mu (t+\pi) \partial^\mu (t + \pi)+1]^2 \ .
\eeq
Only the first term in (\ref{equ:kinetica}) gives rise to the gradient piece $a^{-2}( \partial_i \pi)^2$.  As a result, one requires that 
$M_2^4 \geq 0 $ \cite{Cheung:2007st}, otherwise we could send super-luminal signals using $\pi$.\footnote{Using causality and dispersion relation arguments, one can show $M_2^4 > 0$ in flat space \cite{Adams:2006sv,disps}.} Using this constraint we find
\beq\label{equ:bounda}
\Lambda^4 = 2 M_2^4 + \Mp^2 |\dot H|  > \Mp^2 |\dot H| \ .
\eeq
 
In order to translate (\ref{equ:bounda}) into a useful upper bound on $\Mp^2 |\dot H| $ we must re-write $\Lambda^4$ in terms of known quantities.  Because of the form of the kinetic terms in the action we know $\pi$ scales as $\Lambda^{-2} k^{\tfrac{3}{2} } \omega^{-\tfrac{1}{2}}$. We also know the power spectrum is given by 
\beq\label{equ:generalpower1}
\Delta^2_{\zeta} \simeq \frac{H^4}{4\Lambda^4} c_p^{-3}(\omega \sim H) \sim 10^{-9} \ ,
\eeq
where we used $\zeta = - H \pi$ and that freeze-out occurs at $\omega \sim H$. Here we have assumed there are no large dimensionless parameters floating around. This is clearly not true for the case of excited states with large occupation number, i.e. $\beta_\k \gg 1$, which we cover below. We can now show the following inequality applies:
\beq
\frac{\Mp^2 |\dot H|}{H^4} < \tfrac{1}{4} \Delta_{\zeta}^{-2} c_p^{-3}(\omega \sim H) \ .
\eeq
\vskip 8pt
Now let us return to the case of excited states considered in section \ref{sec:excited1} by modifying (\ref{equ:generalpower1}) to include non-trivial occupation numbers, i.e.
\beq
\Delta^2_{\zeta} \sim \frac{H^4}{4\Lambda^4} c_p^{-3}(\omega \sim H)\Big[ 1+ 2|\beta_{\k_\star}|^2 + 2 \sqrt{|\beta_{\k_\star}|^2 + |\beta_{\k_\star}|^4} \cos(\theta_\k)  \Big] \ ,
\eeq
where we assumed $\beta_\k $ does not vary significantly with $\k$. Therefore, the bound becomes  
\beq
\frac{\Mp^2 |\dot H|}{H^4} < \tfrac{1}{4}\Delta_{\zeta}^{-2}c_p^{-3}(\omega \sim H) \Big[ 1+ 2|\beta_{\k_\star}|^2 + 2 \sqrt{|\beta_{k_\star}|^2 + |\beta_{\k_\star}|^4} \cos(\theta_\k)  \Big] \ .
\eeq

\section{Estimating signal-to-noise for the CMB}\label{append:sn}

In this appendix, we will estimate the signal-to-noise for CMB measurement that are expected from an excited state (in slow-roll).   We will make this estimate by assuming we measure the curvature perturbations directly, rather than temperature fluctuations.  This will serve as an upper bound on the signal-to-noise in the CMB.

\subsection{Signal-to noise estimate in slow-roll models}

If we were to measure the bispectrum of $\zeta$ directly, the signal-to-noise of this measurement would be given by
\beq
\Big(\frac{S}{N} \Big)^2 = \int  \frac{d^3 k_1}{(2\pi)^3}  \frac{3!}{(4\pi^2)}  \int_{1/2}^1 d x_2 \int_{1-x_2}^{x_2} dx_3 x_2^4 x_3^4 B(1,x_2,x_3)^2/ \Delta_\zeta^6 \ .
\eeq
In section \ref{sec:obs}, we showed that excited states with $\beta_{\k_\star} \gtrsim 1$ are ruled out by observations of the power spectrum. Hence, we will work with Bogoliubov states in the limit of small and real $\beta_\k$, such that \cite{Ganc:2011dy}
\bea
B(k_1,k_2,k_3) &\simeq& \beta_{\k_\star}  \frac{1}{2} \frac{H^6}{\Mp^2 \dot \phi^2} \frac{1}{k_1 k_2 k_3}(\frac{1}{k_1^2} + \frac{1}{k_2^2} + \frac{1}{k_3^2} )  \\
&&\times{\rm Re}[ \frac{1- e^{i (k_1+k_2 - k_3)/k_{\star}}}{k_1+k_2 - k_3}+ \frac{1- e^{i (k_1-k_2 +k_3)/k_\star}}{k_1-k_2 + k_3} + \frac{1- e^{i (k_3 +k_2 - k_1)/k_\star}}{k_3 +k_2 - k_1}] \ .
\eea
(Here we have dropped the ${\cal O}(\beta_{\k_\star}^0)$ contribution because it is unobservable.)  

The dominant contribution to the signal-to-noise comes from the flattened limit $k_2 = k_3 - k_1$ (and equivalent configurations).  We may therefore simplify the bispectrum in this folded limit to get
\bea
B(k_1,k_2 \sim k_1 -k_3,k_3) \simeq \beta_{\k_\star}  \frac{1}{2} \frac{H^6}{\Mp^2 \dot \phi^2} \frac{1}{k_1 k_2 k_3^3} {\rm Re}[ \frac{1- e^{i (k_3+k_2 - k_1)/k_{\star}}}{k_3+k_2 - k_1}] \ .
\eea
Evaluating the signal-to-noise analytically is possible, but the full expression is not very insightful. We are most interested in its parametric dependence, which we can understand with the aid of some approximations.  First of all, we notice that the exact folded limit is cut off by the oscillatory term when $x_3\sim1-x_1 + x_\star$, where $x_\star \equiv k_3/ k_\star$.  Using a hard cutoff we can then evaluate the integral over $x_3$, while the $x_2$ integral receives its largest contribution form $x_2 \sim 1$. Therefore we have,
\bea
\Big(\frac{S}{N} \Big)^2  &\sim& \frac{\beta_{\k_\star}^2}{4} \frac{H^{12}}{\Mp^4 \dot \phi^4\Delta_\zeta^6 }\int  \frac{d^3 \k_1}{(2\pi)^3}  \frac{3!}{(4\pi^2)}  \int^{1-x_{\rm max}} dx_2 \frac{1}{ (1-x_2)^2 x_\star} \\
&\sim& 2^4\epsilon^2 \beta^2_{\k_\star} \Delta_\zeta^2  \left(\frac{k_S}{k_L}\right)^2 \frac{ 3!}{(4\pi^2)} {\cal N}_\text{pix},
\eea
where ${\cal N}_\text{pix}$ is the number of independent modes we can measure (i.e the volume of the $k_1$ integral), and we have used $x_\star \sim x_{\rm max}\sim k_L/k_S$ to measure the maximum ratio of momenta where we have the largest contribution to the squeezed limit.

\subsection{Signal-to-noise estimate in small-$c_s$ models}

In small $c_s$ models there is potential increase in the bispectrum due to larger interactions. Specifically we will consider
\beq
H_{\rm int} =\frac{\Mp^2 \dot H}{c_s^2} \int d^3 x a(t)^3 \dot \pi \frac{\partial_i \pi \partial^i \pi}{a^2} \ ,
\eeq
in the limit $c_s\ll 1$, that follows from the EFT of inflation \cite{Cheung:2007st}.  Although the strength of the interaction is enhanced by a factor of $1/c_s^2$, we also expect these will be suppressed in the squeezed limit due to type of derivatively coupled interactions.\\ 

We consider again Bogoliubov states with $\beta_{\k_\star} \ll 1$ due to the power spectrum constraint.  Following the same steps as for the slow-roll case, the dominant contribution to the signal-to-noise comes from the folded limit $k_2 = k_3 - k_1$. We may, as before, simplify the bispectrum in this limit and get
\bea
B(k_1,k_2,k_3) \sim \frac{\beta_{\k_\star} }{4 c_s^2} \Delta^4_\zeta \frac{(k_1^2(k_1^2 -k_2^2 -k_3^2)+{\rm permutations})}{k_1^3 k_2^3 k_3^3}{\rm Re}[ \frac{1- e^{i (k_3+k_2 - k_1)/k_{\star}}}{k_3+k_2 - k_1}]  \ ,
\eea
where we used that, $c_s \ll 1$, the mode functions for $\zeta$ take the form
\beq
\zeta_\k = \frac{H^2}{2 \Mp |\dot H|^{1/2}} \frac{1}{k^{3/2}} (1- i c_s k \tau) e^{i k c_s \tau} \ .
\eeq
Again we wish to calculate the signal-to-noise:
\bea
\Big(\frac{S}{N} \Big)^2 &\simeq&  \frac{\beta_{\k_\star}^2 }{16 c_s^4}\Delta^2_\zeta  \int  \frac{d^3 \k_1}{(2\pi)^3}  \frac{3!}{(4\pi^2)} \\&&\times  \int_{1/2}^1 d x_2 \int_{1-x_2+x_\star}^{x_2} dx_3 [\tfrac{(1-x_2^2-x_3^2)+x_2^2 (x_2^2-x_3^2 -1)+x_3^2(x_3^2-x_2^2-1)}{(x_3+x_2 - 1)x_2 x_3}]^2   \ ,
\eea
and we find 
\bea
\Big(\frac{S}{N} \Big)^2 &\simeq&  \frac{\beta_{\k_\star}^2 }{16 c_s^4}\Delta^2_\zeta  \int  \frac{d^3 \k_1}{(2\pi)^3}  \frac{3!}{(4\pi^2)} \times  \int_{1/2}^1 d x_2\frac{(1-x_2)^2}{x_\star}  \ .
\eea
Notice we do see an enhancing factor of ${x_\star}^{-1}$, however it is easy to see that this integral does not get contributions from $x_2 \sim 1$ (or equivalently $x_3 \to 0$).  As a result, there is no improvement for the signal-to-noise in the squeezed limit for small speed of sound models.

%\begingroup\raggedright\begin{thebibliography}{10}

%\endgroup

\end{document}